\documentclass[epj,twocolumn]{webofc}
\usepackage[varg]{txfonts}   
\woctitle{ISVHECRI 2014}
\begin{document}
\title{Calculation of conventional and prompt lepton fluxes at very high energy}
\author{Anatoli Fedynitch\inst{1,2}\fnsep\thanks{\email{anatoli.fedynitch@cern.ch}} 
      \and Ralph Engel\inst{1}\fnsep \and
        Thomas K. Gaisser\inst{3}\fnsep \and
        Felix Riehn\inst{1}\fnsep \and
        Todor Stanev\inst{3}
}

\institute{Karlsruher Institut f\"ur Technologie, Institut f\"ur Kernphysik, Postfach 3640, 76021 Karlsruhe, Germany
\and
           CERN EN-STI-EET, CH-1211 Geneva 23, Switzerland
\and
           Bartol Research Institute, Department of Physics and Astronomy, University of Delaware, Newark, DE 19716, USA
          }

\abstract{%
  An efficient method for calculating inclusive conventional and prompt atmospheric leptons fluxes is presented. The coupled cascade equations are solved numerically by formulating them as matrix equation.
  The presented approach is very flexible and allows the use of different hadronic interaction models,
  realistic parametrizations of the primary cosmic-ray flux and the Earth's atmosphere, and a detailed treatment of particle interactions and decays. The power of the developed method is illustrated by 
  calculating lepton flux predictions for a number of different scenarios.
}
\maketitle
\renewcommand{\thefootnote}{\fnsymbol{footnote}}
\section{Introduction}
\label{sec:intro}

Cosmic rays entering the Earth's atmosphere produce a multitude of secondary particles in interactions with air nuclei. Some of the secondary particles decay into muons and neutrinos, which are not absorbed in the atmosphere and can reach particle detectors at ground level. The spectra of these leptons contains not only information about the primary cosmic rays, but also about the particle physics of their production and the properties of the traversed atmosphere. Furthermore, searches for high-energy neutrinos from astrophysical sources have to cope with a large flux of atmospheric leptons as background. A better understanding of this flux, in particular its dependence on zenith angle and the changing properties of the atmosphere will help to develop improved methods to identify astrophysical neutrino fluxes and also contribute to a better understanding of hadronic interactions at high energy.

Many calculations of atmospheric lepton fluxes have been carried out since the early 1960's (e.g.~\cite{Zatsepin:1962ta}, see \cite{Gaisser:2002jj} for a review). However, most of them incorporated approximations in solving the cascade equations that lead to increased uncertainties on the relation between the predicted fluxes and the physical input parameters, or the calculations could only be carried out in detail for just one or a few parameter/model combinations due to the large CPU time requirements. In this work we reduce the uncertainties related to the calculation method to a minimum by developing a numerical method with a level of detail comparable with Monte Carlo calculations \cite{HKKM_2011,bartol_2004,Fedynitch:2012ii}. 
In addition, the contributions of heavy flavor mesons and resonances to the flux of atmospheric leptons is accounted for in detail. The code is made publicly available\footnote[1]{https://github.com/afedynitch/MCEq}.

Our approach is based on the numerical solution of the coupled cascade equations, which have been rewritten into a matrix form to make use of modern implementations of linear algebra algorithms.
While providing superior precision at very high energies where Monte Carlo methods are often statistically inefficient, the high performance of the algorithm allows us to perform calculations for many input parameter and model assumptions in a very short time.
On an average portable computer it takes a few seconds to calculate lepton fluxes, while keeping most of relevant parameters accessible for users and easy to modify according to the current application.

\section{Coupled cascade equation}
\label{sec:cce}

The cascade equations for particle $h$ can be written for one discrete energy bin $E_i$
\begin{subequations}
\begin{align}
\frac{{\rm d}\Phi^h_{E_i}}{{\rm d}X}  =
\label{eqn:sink_int} &-\frac{\Phi^h_{E_i}}{\lambda^h_{int, E_i}}\\
\label{eqn:sink_dec} &-\frac{\Phi^h_{E_i}}{\lambda_{dec,E_i}^h(X)}\\
\label{eqn:source_int} & + \sum_{E_k \ge E_i}{\sum_l{\frac{c_{l(E_k) \to h(E_i)}}{\lambda^l_{int,E_k}}\Phi^l_{E_k}}}\\
\label{eqn:source_dec} & + \sum_{E_k \ge E_i}{\sum_l{\frac{d_{l(E_k) \to h(E_i)}}{\lambda^l_{dec,E_k}(X)}\Phi^l_{E_k}}}.
\end{align}
\end{subequations}
 It is part of a system of coupled ordinary differential equations, describing the evolution of the flux $\Phi$ of particles 
 as a function of the atmospheric slant depth
\begin{equation}
\label{eq:slant_depth}
  X(h_O) = \int_0^{h_O}~{\rm d}l~\rho_{air}(h_{atm}(l)).
\end{equation}
For an observation height $h_O$, $X(h_O)$ is computed along the trajectory $l$ of the cascade core through the atmosphere using the mass density $\rho$, which is typically a function of the atmospheric height. The behavior of the particle cascade is driven by the competition of two source terms (\ref{eqn:source_int}), (\ref{eqn:source_dec}) and two sink terms (\ref{eqn:sink_int}), (\ref{eqn:sink_dec}). The interaction length $\lambda^h_{int,E_i} = m_{air}/\sigma^{\rm inel}_{p-air}(E_i)$ in units g/cm$^2$ \cite{gaisser_book} is independent of the slant depth and only varies slowly with energy due to the inelastic particle-air cross-section. The decay length $\lambda^h_{dec,E_i}(X) = c\tau_h E_i \rho_{air}(X)/m_h$ is proportional to the life-time $\tau_h$ of particle $h$ and can vary by orders of magnitude due to the relativistic time dilation. In our approximation new particles are created along the shower trajectory in hadronic interactions in (\ref{eqn:source_int}) or decays in (\ref{eqn:source_dec}), with energy conservation restricting the range of possible source particles to energies $E_k$ equal or greater than $E_i$.

The interactions coefficients of particle $l$ producing particle $h$, $c_{l(E_k) \to h(E_i)}$, are obtained from hadronic interactions models by histogramming the particle yield as function of $x^{(p)}_{lab} = E_i/E_k$ for $l$-air collisions. Suitable interaction models include, for example, {\sc QGSJET-II-04} \cite{Ostapchenko:2011bda} and {\sc EPOS LHC} \cite{epos}, and the upcoming versions of {\sc SIBYLL} \cite{sibyll_1994} and {\sc DPMJET-III} \cite{dpmjetIII}. Alternatively one can directly extrapolate results of fixed-target experiments on light nuclear targets using scaling arguments.

While in some cases it is possible to find analytical expressions for the decay coefficients $d_{l(E_k) \to h(E_i)}$, see ~\cite{gaisser_book}, numerical simulations have to be used for describing complex decays accurately. We have tabulated the decay coefficients as function of $E_k$ based on simulations with the Monte Carlo event generator {\sc Pythia 8}~\cite{pythia,Sjostrand:2008bk} that includes also rare decay channels and accounts for the effect of electroweak matrix elements where applicable.

The full system of equations of the hadronic cascade can be obtained by writing Eq.~(1) for all possible types of hadrons and leptons. In the following we will concentrate on high lepton energies and neglect the interaction and/or decay terms of neutrinos and muons. To obtain, for example, the flux of atmospheric neutrinos at the surface, one needs to solve the full system taking into account the non-linear $X$ dependence of $\lambda_{dec}$ and the non-analytic forms of the particle spectra serving as input for the interaction coefficients $c_{l \to h}$.

\subsection{Matrix form}
\label{ssec:matrix_form}

An efficient numerical computing scheme can be found by rewriting the cascade equations into matrix form. We group the different $\Phi_h(E_i)$ into a column vector $\Phi$ by writing blocks for each particle type for the discrete energy spectra
\begin{equation}
\label{eq:state_vector}   
 {\Phi}=\begin{pmatrix} \Phi_p(E_0) \\
 			\Phi_p(E_1) \\
          \cdots \\ 
          \Phi_p(E_N) \\[2mm]
            \Phi_n(E_0) \\ 
            \cdots
        \end{pmatrix}.
\end{equation}
The energy grid $E_i = 50\,{\rm GeV}\cdot 10^{i/N}$ 
is logarithmically spaced with roughly 8 bins per decade of energy across the energy range of the calculation between $50$\,GeV and $10^{10}$\,GeV. We explicitly include more than 50 types of mesons, baryons and leptons. Together with some additional technical groups the dimension of $\Phi$ is then $\sim 6000$.

The reciprocal coefficients $1/\lambda$ of interaction and decay lengths are arranged in diagonal matrices
\begin{align}
\nonumber \boldsymbol{\Lambda_{int}} = {\rm diag}(&\frac{1}{\lambda^p_{int,E_0}} \cdots \frac{1}{\lambda^p_{int,E_N}},\\ 
&\frac{1}{\lambda^n_{int,E_0}}, \cdots,  \frac{1}{\lambda^n_{int,E_N}},\\
&\frac{1}{\lambda^{\pi^+}_{int,E_0}}, \cdots ).
\end{align}
The decay length matrix $\boldsymbol{\Lambda_{dec}}$ is constructed analogously using $\tilde{\lambda}^h_{dec,E_i} = \lambda^h_{dec,E_i}(X)/\rho_{air}(X)$ to factorize out the dependence on the air density. Sub-matrices containing the interaction coefficients are defined as
\begin{equation}
\boldsymbol{C}_{l \to h} =
          \begin{pmatrix} c_{l(E_0) \to h(E_0)} &\cdots  & c_{l(E_0) \to h(E_N)} \\ 
          &&c_{l(E_1) \to h(E_N)}\\
          &  \ddots  & \vdots\\
          \mbox{\Large 0} &   &c_{l(E_N) \to h(E_N)} \\ \end{pmatrix},
\end{equation}
and sub-matrices for the decay $\boldsymbol{D}_{l \to h}$ are constructed in a similar way. The full interaction and decay matrices $\boldsymbol{C}$ and $\boldsymbol{D}$ are built from these sub-matrices according to the order of particle types in $\Phi$ 
\begin{equation}
\label{eq:C_matrix} \boldsymbol{C} =
          \begin{pmatrix} \boldsymbol{C}_{p \to p} & \boldsymbol{C}_{n \to p} &\boldsymbol{C}_{\pi^+ \to p} & \cdots   \\ 
          \boldsymbol{C}_{p \to n} & \boldsymbol{C}_{n \to n} & \boldsymbol{C}_{\pi^+ \to n}& \cdots\\
          \boldsymbol{C}_{p \to \pi^+} & \boldsymbol{C}_{n \to \pi^+} & \boldsymbol{C}_{\pi^+ \to \pi^+}& \cdots\\
          \vdots & \vdots & \vdots & \ddots\end{pmatrix}.
\end{equation}
Using the definitions above, the matrix form of the coupled cascade equations can be written as
\begin{equation}
\label{eq:matrix_ce}
  \frac{\rm{d}}{\mathrm{d}X}\boldsymbol{\Phi} = \left[(-\boldsymbol{1} + \boldsymbol{C}){\boldsymbol{\Lambda}}_{int} + \frac{1}{\rho(X)}(-\boldsymbol{1} + \boldsymbol{D})\boldsymbol{\Lambda}_{dec}\right]\boldsymbol{\Phi}.
\end{equation}

\subsection{Short-lived particles}

\label{ssec:short_lived}
One goal of this work is to accurately take into account contributions of heavy flavor mesons and resonances to the flux of atmospheric leptons. Their short decay lengths introduce quickly decaying modes in Eqs.~(\ref{eqn:sink_dec}) and (\ref{eqn:source_dec}), which appear as eigenvalues of the decay matrix $\boldsymbol{D}$ with a large modulus of the negative real part.
\begin{figure}
  \centering
  \includegraphics[width=\columnwidth]{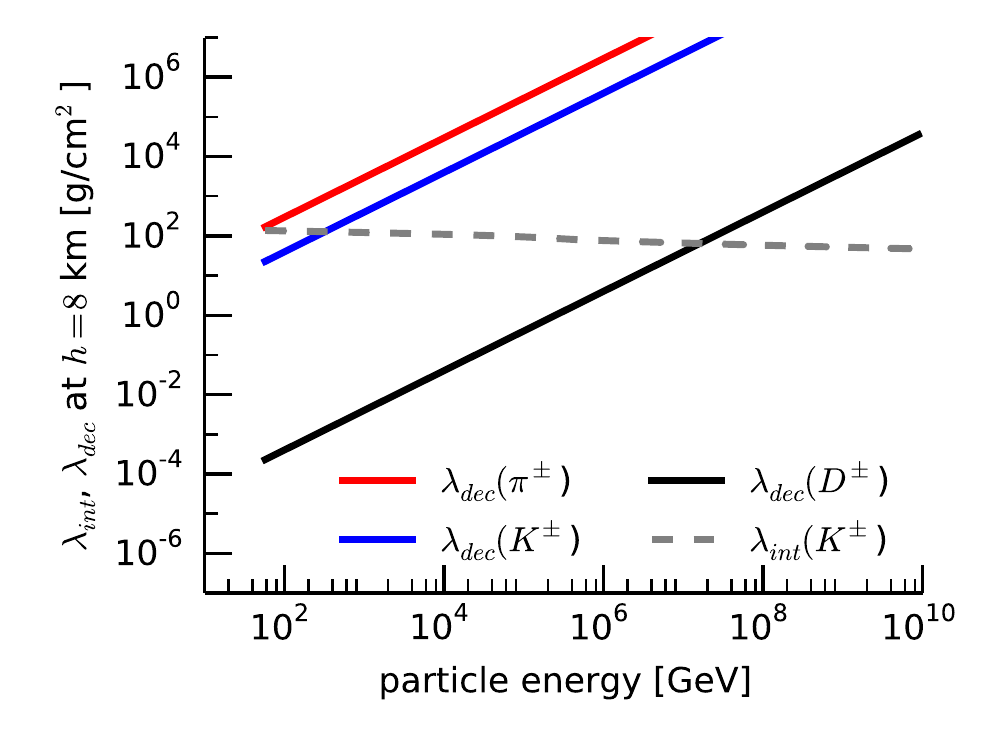}
  \caption{Decay lengths $\lambda_{dec}$ for a subset of hadrons, evaluated at $h_{atm}=8$ km. Superimposed is the interaction length $\lambda_{int}$ of $K^\pm$.\label{fig:int_decay_lengths}}
\end{figure}
Fig.~\ref{fig:int_decay_lengths} illustrates the large difference between the decay lengths of conventional and charmed mesons. The energy dependence originates from time dilation. If only pions and kaons would be considered as intermediate mesons, the cascade equations would become interaction dominated above 1\,TeV. The decay would be a slow process and the equations could be easily integrated using a moderate step size of $\mathcal{O}(1~{\rm g/cm}^2)$. If short-lived particles are included, the choice of the appropriate step size is driven by the smallest eigenvalue, avoiding oscillations of fluxes around zero that are unphysical ($\Phi \ge \boldsymbol{0}$). For this reasons the equation system becomes a stiff numerical problem involving step sizes of $\mathcal{O}(10^{-5}\,{\rm g/cm}^2)$. From a performance point of view this approach is unreasonable since the integration can run up to values of $X \sim \mathcal{O}(10^4)$ g/cm$^2$.

To reduce the stiffness of the equation system, we introduce the resonance approximation. In the resonance approximation very short-lived particles (resonances), e.g.\ $\eta$ or $\rho$ mesons, decay immediately after their creation at the vertex. For each particle $h$ this approximation is valid in a regime, where 
\begin{equation}
\label{eq:resonance_condition}
\lambda^h_{dec} \ll \lambda^h_{int}.
\end{equation}
The coefficients for the chained production of a long lived secondary particle $l$ via production and immediate decay of the resonance $\eta$, neglecting its interactions, can be written as
\begin{equation}
\label{eq:chained_decay}
\boldsymbol{R}_{h \to \eta \to l} = \boldsymbol{D}_{\eta \to h} \cdot \boldsymbol{C}_{h \to \eta}.
\end{equation}
To understand this result, let $\boldsymbol{\eta}^{int}_n$ be the vector containing the fluxes of all resonances, which are created during the integration step $n$. Using the matrix notation and forward Euler integration we can write
\begin{equation}
\label{eq:res_creation}
\boldsymbol{\eta}^{int}_{n} = \boldsymbol{C}^{res}_{h \to \eta}\boldsymbol{\Lambda_{int}}~\Phi \cdot \Delta X_n.
\end{equation}
$\boldsymbol{C}^{res}_{h \to \eta}$ denotes a ($k \times d_\Phi$) matrix similar to $\boldsymbol{C}$ defined in Eq.~(\ref{eq:C_matrix}), which contains production coefficients for resonances in interactions of hadrons. According to the approximation, all created resonances have to decay into ordinary particles within the same integration step. By writing this condition for a single resonance type $k$
\begin{equation}
\nonumber \eta_{k,n+1} \equiv 0
=  \eta_{k,n}  - \frac{1}{\lambda^{\eta_k}_{dec,eff}} \eta_{k,n} \cdot \Delta X_{n},
\end{equation}
we obtain the effective decay length
\begin{equation}
\label{eqn:eff_decay}
\lambda^{\eta_k}_{dec,eff} = \Delta X_{n} = \rho(X) \widetilde{\lambda}^{\eta_k}_{dec,eff}.
\end{equation}
Using ${\lambda^{\eta_k}_{dec,eff}}$ instead of the true decay length for short-lived resonances we make sure that all particles decay after one integration step in $X$ without having to change the numerical treatment of the cascade equations. 
It should be noted that ${\lambda^{\eta_k}_{dec,eff}}$ does not depend on the properties of the resonance. The contribution to the ordinary particle flux due to resonance decay is then
\begin{align}
\nonumber \Delta \Phi^{res}_{n+1} &= \boldsymbol{D}^{res}_{\eta \to h}\boldsymbol{\Lambda}^{res}_{dec,eff}~\boldsymbol{\eta} \cdot \frac{\Delta X_{n}}{\rho(X)}\\
\label{eq:res_decay} &=\boldsymbol{D}^{res}_{\eta \to h}~\boldsymbol{\eta},
\end{align}
where $\boldsymbol{D}^{res}_{\eta \to h}$ is a ($d_\Phi \times k$) matrix, containing decay coefficients of resonances into hadrons and leptons. 
By inserting Eq.~(\ref{eq:res_creation}) in (\ref{eq:res_decay}) we replicate the expression from Eq.\ (\ref{eq:chained_decay}) for the production of particles via intermediate resonances
\begin{align}
\nonumber \Delta \Phi^{\to \eta \to}_{n+1} &= (\boldsymbol{D}^{res}_{\eta \to h} \cdot \boldsymbol{C}^{res}_{h \to \eta})\boldsymbol{\Lambda}_{int}~\Phi_n \cdot \Delta X_n\\
&=\boldsymbol{R}~\boldsymbol{\Lambda}_{int}~\Phi_n \cdot \Delta X_n
\end{align}
and define the square ($d_\phi \times d_\phi$) resonance matrix $\boldsymbol{R}$. Chained decays proceeding through two or more resonances are governed by additional left multiplications of decay matrices. Extending the matrix form of the cascade equations (\ref{eq:matrix_ce}) with intermediate resonance production results in
\begin{align}
\label{eq:matrix_ce_res}
\nonumber \frac{\rm{d}}{\mathrm{d}X}\boldsymbol{\phi} = &(-\boldsymbol{1} + \boldsymbol{C} + \boldsymbol{R}){\boldsymbol{\Lambda}}_{int}~\Phi\\
+ \frac{1}{\rho(X)}&(-\boldsymbol{1} + \boldsymbol{D})\boldsymbol{\Lambda}_{dec}~\Phi.
\end{align}
At high energies, where
\begin{equation}
   \lambda_{dec} \approx \lambda_{int},
\end{equation}
the interaction of resonances becomes important. We introduce the parameter $t_{mix} = \lambda_{dec}(E)/\lambda_{int}(E)$, which is a threshold value, separating the energy regime where the particle can be treated as resonance from a regime where it has to be a full member of the cascade and listed in $\Phi$. A reasonable value is $t_{mix} = 0.05$. This complicates somewhat the procedure to fill the $\boldsymbol{C}$, $\boldsymbol{D}$ and $\boldsymbol{R}$ matrices, where for each particle the individual threshold has to be taken into account as cut in row and/or column.

\section{Calculation input}

\subsection{Initial state}
\label{ssec:primary_flux}

The initial state of the cascade equation is the flux of cosmic rays at the top of the atmosphere. Since the calculation relies on the properties of the average air shower, the superposition theorem is sufficient to model the flux and composition. A cosmic ray nucleus is modeled as $Z$ protons and $A-Z$ neutrons, with each nucleon carrying the fraction $E/A$ of the total kinetic energy. The relevant input for the initial condition is, therefore, the all-nucleon spectrum separated in the proton and neutron components. Alternatively one can use a single particle per energy bin and calculate particle yields at the surface for comparisons with full Monte Carlo methods such as \cite{HKKM_2011,bartol_2004,Fedynitch:2012ii}.
\begin{figure}[htb]
  \centering
  \includegraphics[width=0.95\columnwidth]{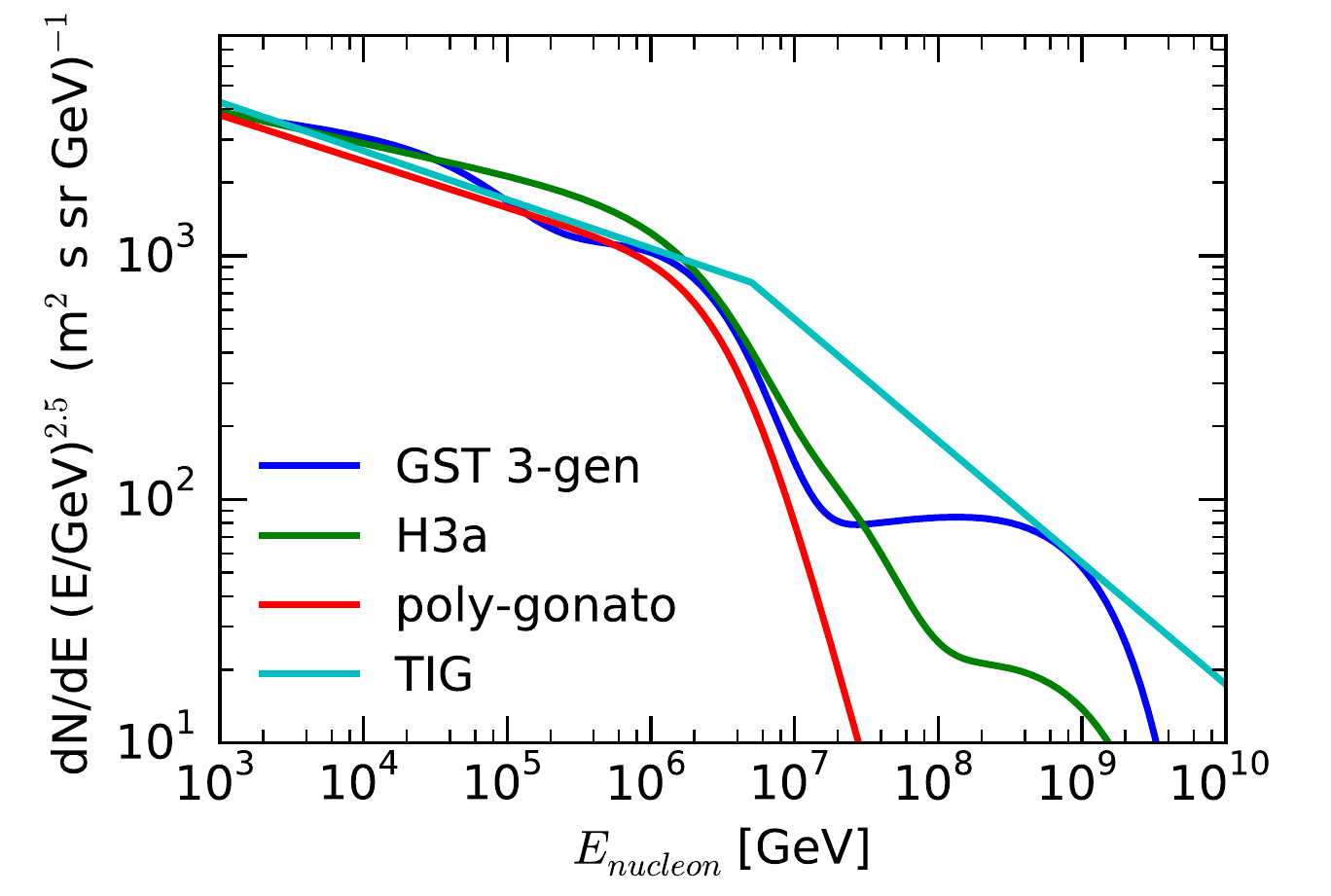}
  \includegraphics[width=0.95\columnwidth]{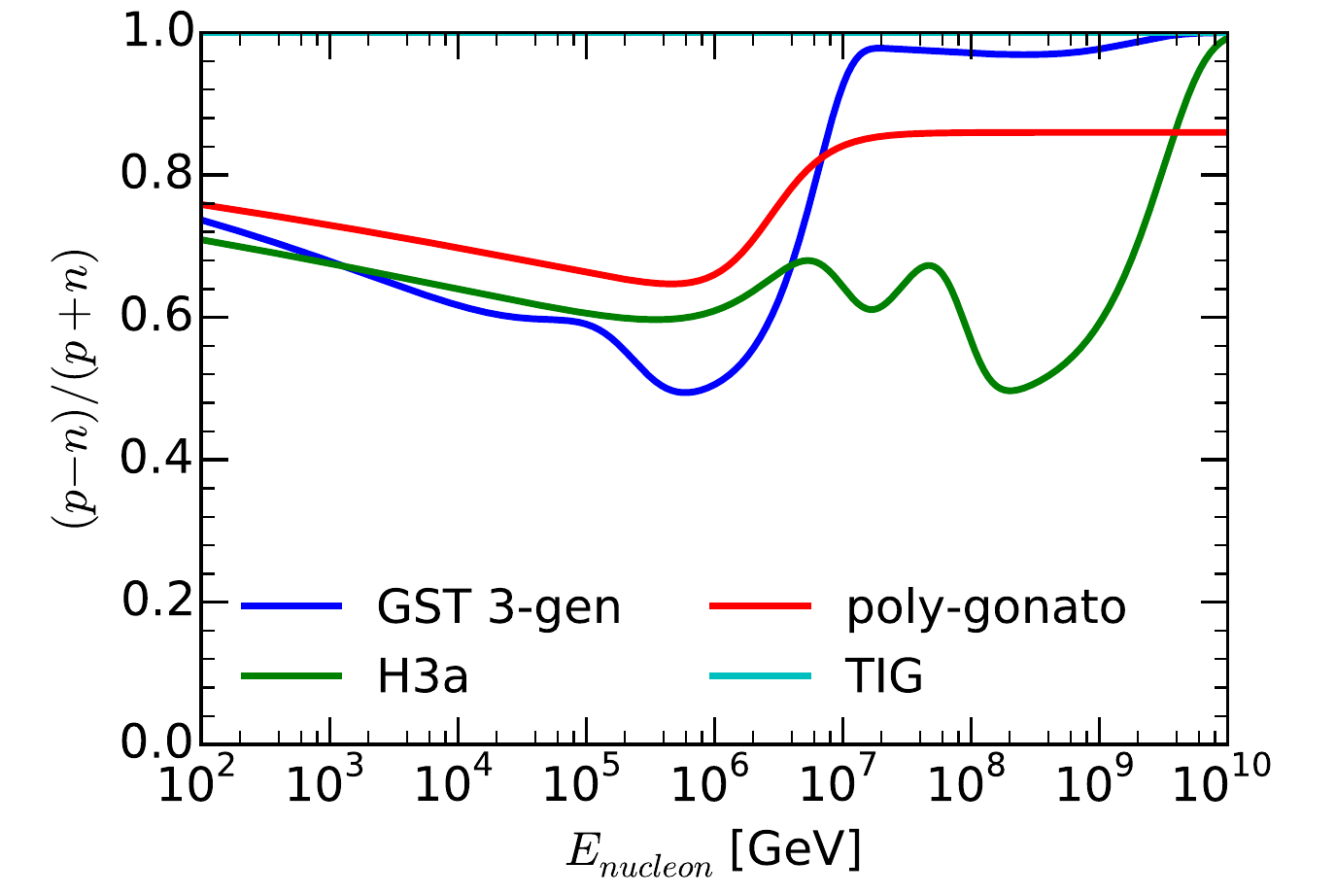}
  \caption{Models of the cosmic ray nucleon spectrum and the neutron fraction. Gaisser-Stanev-Tilav (GST)~\cite{Gaisser:2013tu} and Hillas-Gaisser (H3a)~\cite{Gaisser:2012em} are recent 3 generation/5 mass component models. The proton-only broken power law model by Thunman et al.\ (TIG)~\cite{thunman_1996} has been often used for calculation of the prompt flux in the past. The poly-gonato model~\cite{Horandel:2003go} focuses on the flux below and at the knee and it is not applicable at very high energies.
  \label{fig:primary_spectrum}}
\end{figure}
\begin{figure}
  \centering
  \includegraphics[width=0.95\columnwidth]{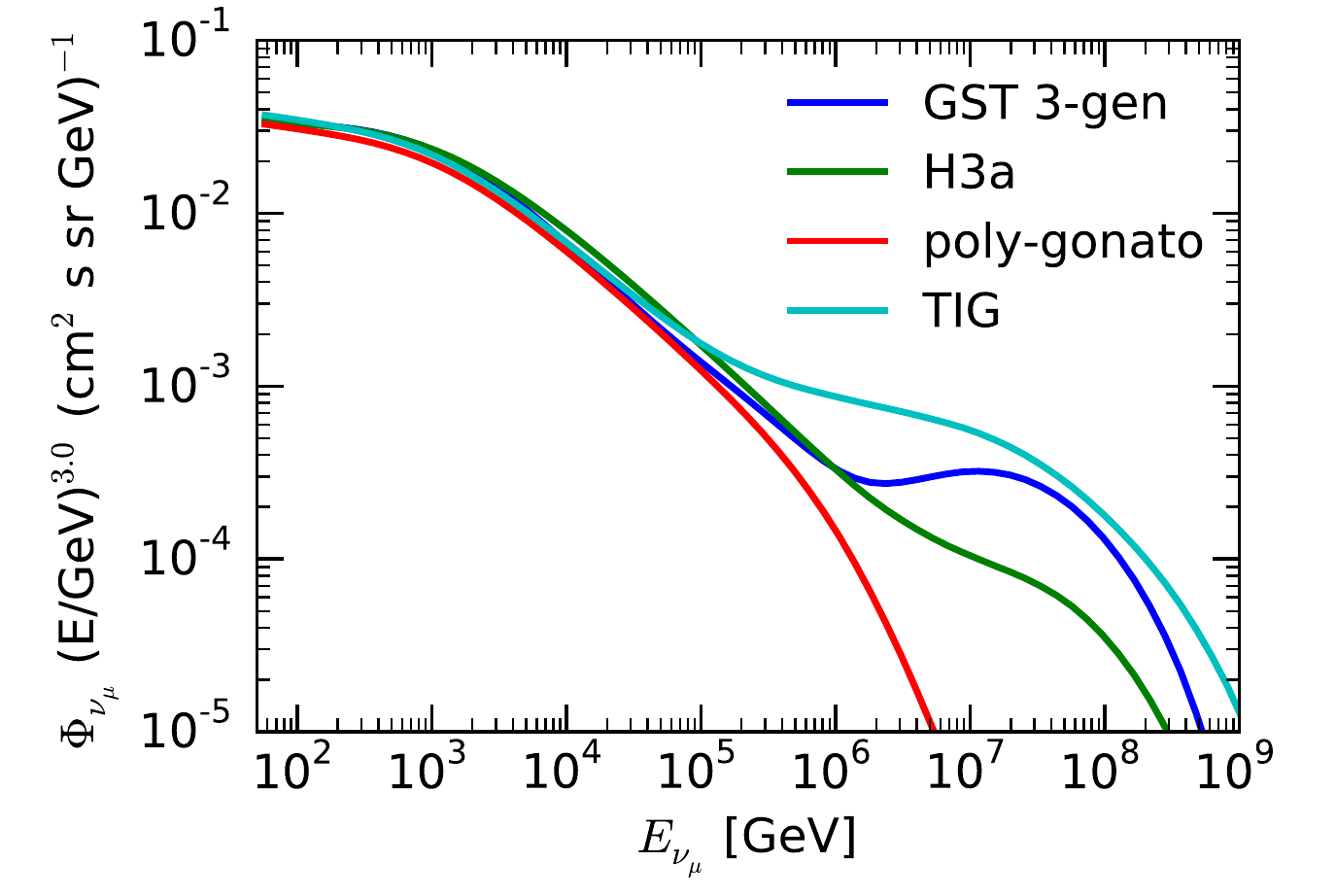}
  \caption{Primary model dependence of the atmospheric conventional + prompt neutrino flux. The model abbreviations are described in the caption of Fig.~\ref{fig:primary_spectrum}.
  \label{fig:primary_variation}}
\end{figure}
Fig.~\ref{fig:primary_spectrum} shows the spectra of three models we typically use for calculations. We emphasize, that it is crucial to include the knee and ankle in the flux calculations with respect to the range of energies accessible by current neutrino observatories, such as IceCube and Antares. In Fig.~\ref{fig:primary_variation} the influence of the primary model on the neutrino flux is shown. While at energies below the knee, where direct measurements of the cosmic ray flux are available, the difference between the models is small, there are large uncertainties at tens of PeV. Improving the knowledge of the spectrum and composition as measured by air shower experiments would help to disentangle these ambiguities.

\subsection{Geometry and Atmosphere}
\label{ssec:geometry}
Treating the atmosphere in planar approximation, the relation between the height, slant depth, and local density can be taken directly from measurements.
An often used approach, based on the idea by Linsley \cite{CORSIKA_report}, is a parametrization of the relation between height and mass overburden $X_v(h)$ (slant depth for vertical trajectory) using 5 piecewise defined exponential functions, representing layers of the atmosphere. A higher flexibility is achieved if tabulated atmospheric data
(e.g.\ from satellites) or detailed numerical models, such as NRLMSISE-00 \cite{Picone:2002go}, are used. At large zenith angles also the curvature of the surface of the Earth has to be accounted for. Therefore we compute and tabulate the relation $\rho(h_{atm}(X))$ for each provided zenith angle $\theta$ and parametrization of the atmosphere.
\begin{figure}
  \centering
  \includegraphics[width=0.95\columnwidth]{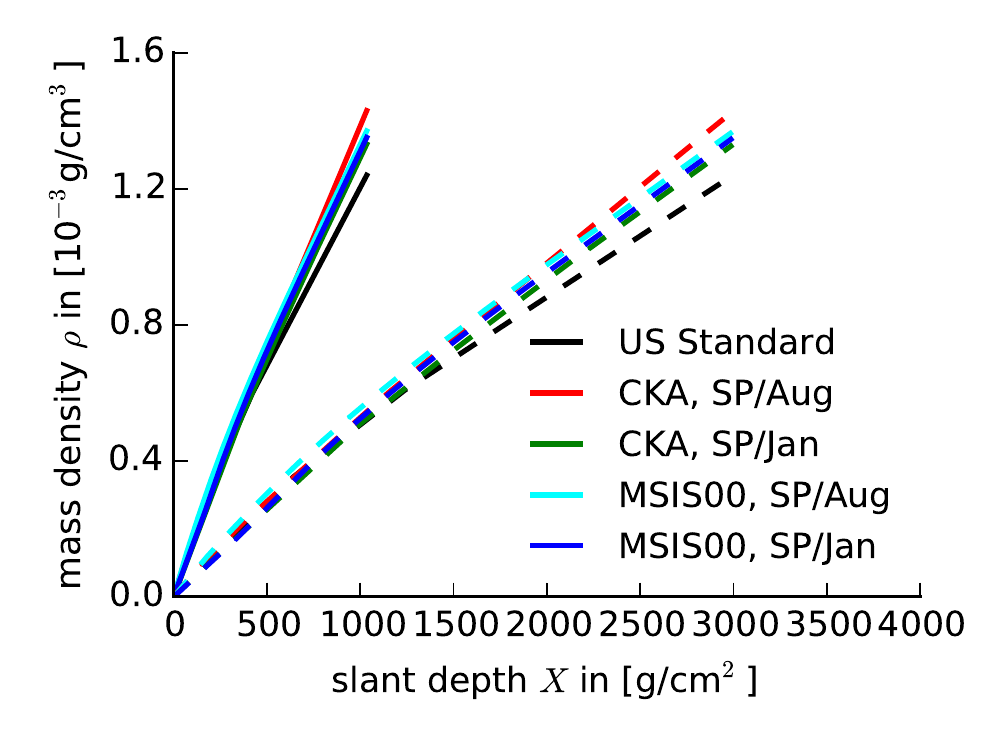}
  \caption{Atmospheric density dependence on $X$, calculated using parameterizations for the US Standard Atmosphere~\cite{us_std_atmosphere} and the South Pole as implemented in CORSIKA~\cite{Heck:ut}, and the NRLMSISE-00 model. Solid lines represent a trajectory for $\theta = 0^\circ$ and dashed for $\theta = 70^\circ$.
  \label{fig:atm_comparison}}
\end{figure} 
As shown in Fig.~\ref{fig:atm_comparison}, this results in a linear smooth curve which is well suited for interpolation with splines.
\begin{figure*}
  \centering
  \includegraphics[width=0.975\textwidth]{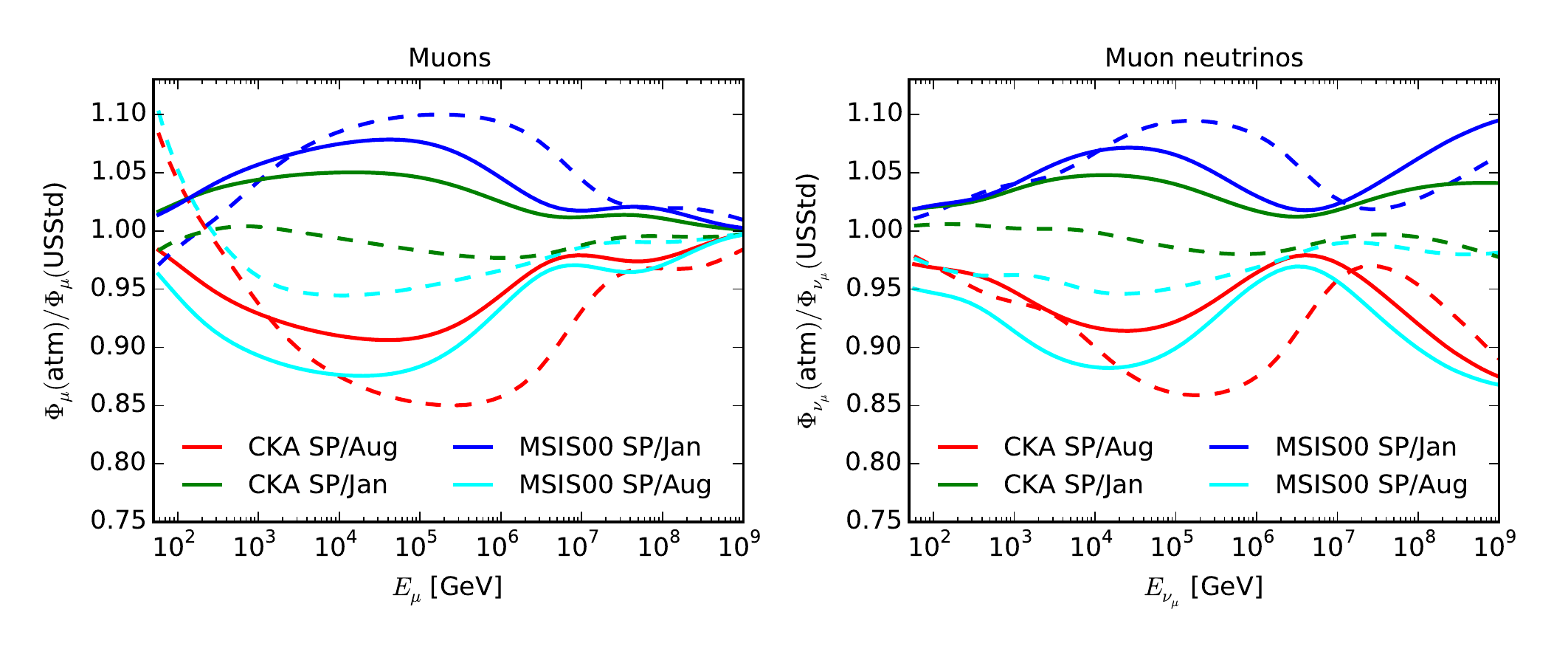}
  \caption{Ratio of the flux calculated with different atmospheric models to the flux with US Standard atmosphere (USStd). The parameters and names are described in the caption of Fig.~\ref{fig:atm_comparison}. The primary model is H3a and the interaction model {\sc SIBYLL-2.3 RC1}. A vertical trajectory ($\theta = 0^\circ$) is represented by solid and a horizontal ($\theta = 90^\circ$) by dashed lines.
  \label{fig:atm_flux}}
\end{figure*}

In Fig.~\ref{fig:atm_flux} the ratio of the flux $\Phi$, calculated with different models of the atmosphere to the flux calculated using the US Standard Atmosphere~\cite{us_std_atmosphere} is shown. Seasonal variations are of comparable magnitude in the CORSIKA parameterizations as in NRLSMSISE-00. Both models predict a seasonal variation of muon and neutrino rates at the order of $\pm 10\%$ in agreement with what IceCube has observed\cite{Desiati:2011bg,Desiati:AEc-X2-i}. Another important feature is the non-trivial zenith angle dependence on atmospheric variations. A more detailed modeling of the atmosphere can be used to improve the experimental investigation of the prompt flux \cite{Desiati:2010wt}. Furthermore, the flux at energies $>$ PeV, which is dominated by prompt leptons, exhibits also a $\sim 15\%$ dependence on the atmosphere.

\section{Applications}
\subsection{Calculation of the prompt flux}
\label{ssec:charm_models}
In a related contribution \cite{Riehn:ISVHECRI}, we discuss a model of charmed hadron production as it is implemented in the Monte Carlo model {\sc SIBYLL-2.3 RC1}. For the calculation of the prompt flux using the method of this work, it is sufficient to take the interaction cross sections and 
Feynman-$x_F$ or the $x_{Lab} = E_{\rm secondary}/E_{\rm projectile}$ distributions from Monte Carlo. Two alternative models, where it is possible to extract $x_{Lab}$ distributions, are the Martin-Ryskin-Stasto (MRS) \cite{2003AcPPB..34.3273M} model and the charm model of {\sc DPMJET-II.55} \cite{Berghaus:2008ed}. MRS considers perturbative production of charm quarks, based on a saturation model, and {\sc DPMJET} includes contributions from non-perturbative, perturbative and fragmentation mechanisms, similar to {\sc SIBYLL-2.3}.
\begin{figure}
  \centering
  \includegraphics[width=0.91\columnwidth]{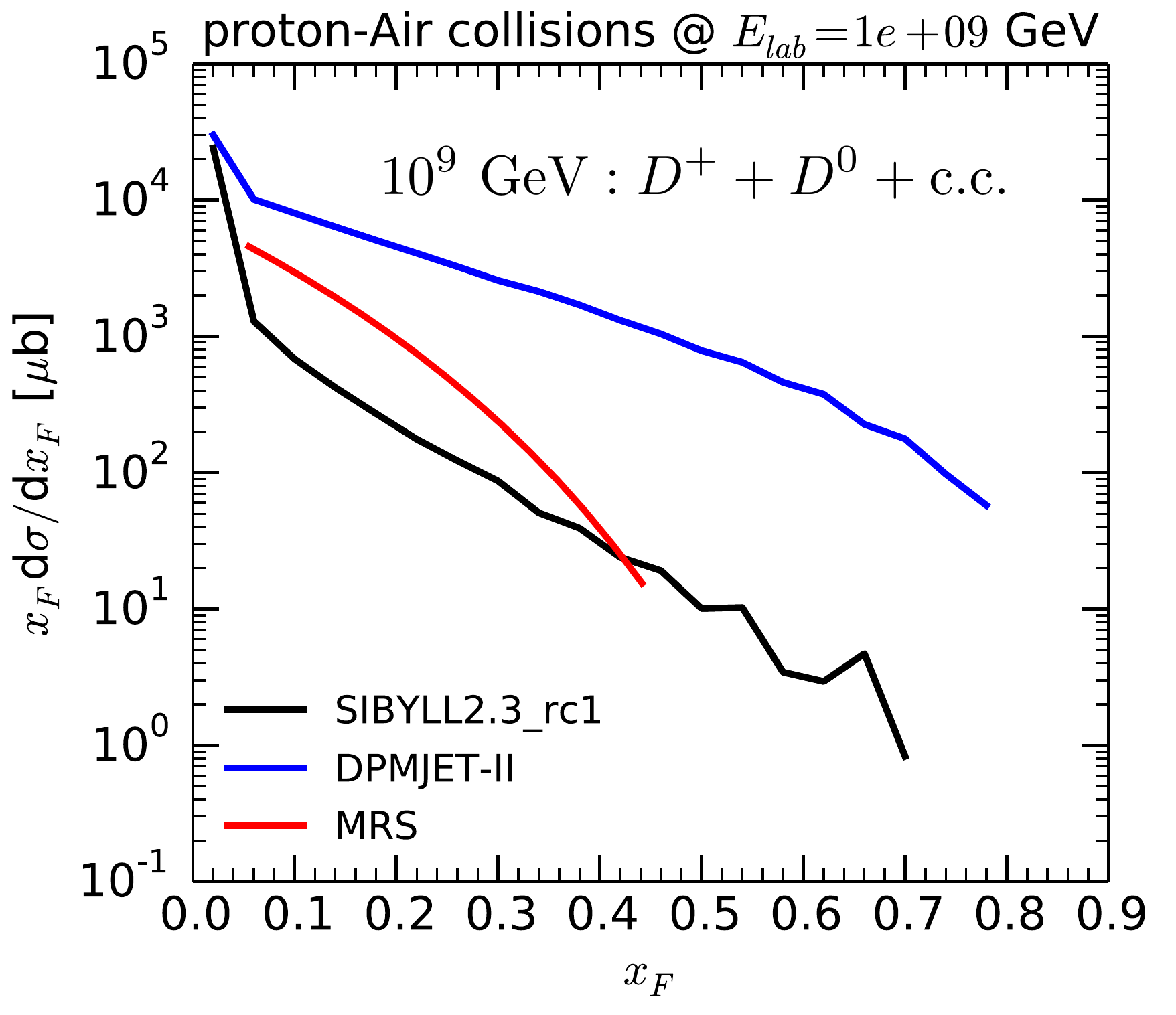}
  \includegraphics[width=0.91\columnwidth]{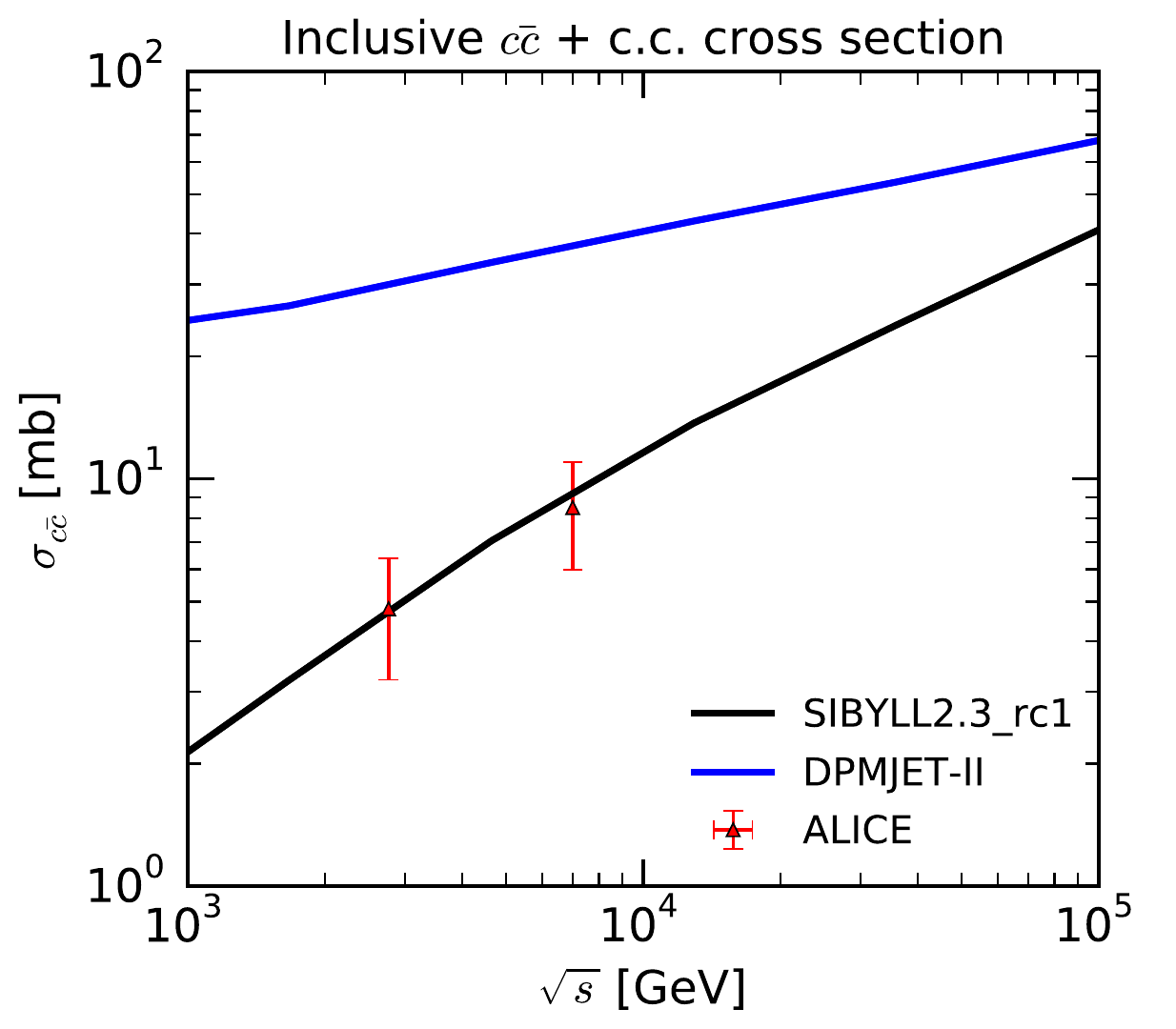}
  \caption{(top) Feynman-$x_F$ distributions as predicted by {\sc SIBYLL-2.3 RC1}, the MRS model and DPMJET-II. (bottom) Inclusive $c\bar{c}$ cross-section in $pp$ collsions. The ALICE measurement is corrected for the invisible part of the cross-section and extrapolated to full phase-space \cite{2012JHEP...07..191A}.
  \label{fig:charm_models}}
\end{figure}

The comparison in Fig.~\ref{fig:charm_models} shows large differences in shape and cross-section between the models. MRS predicts a softer spectrum with a moderate forward cross-section. Due to a hard spectrum and the largest cross-section, we can expect that calculations using {\sc DPMJET} will result in the highest fluxes. {\sc SIBYLL} should produce more inclusive leptons compared to MRS, since the harder spectrum will yield charmed mesons at higher $x$ (see discussion of spectrum weighted moments in \cite{Riehn:ISVHECRI}). 
\begin{figure}
  \centering
  \includegraphics[width=0.925\columnwidth]{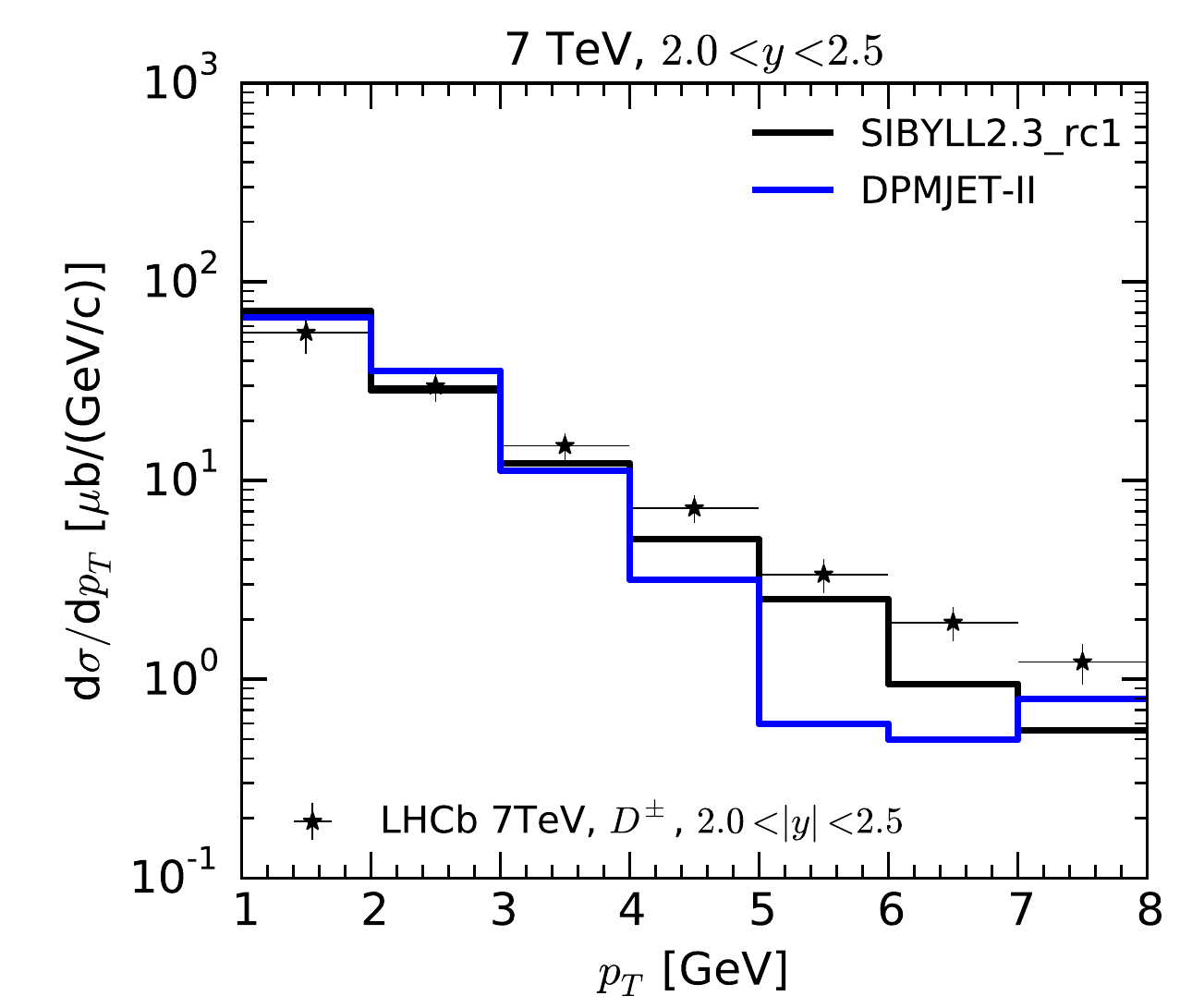}
  \includegraphics[width=0.925\columnwidth]{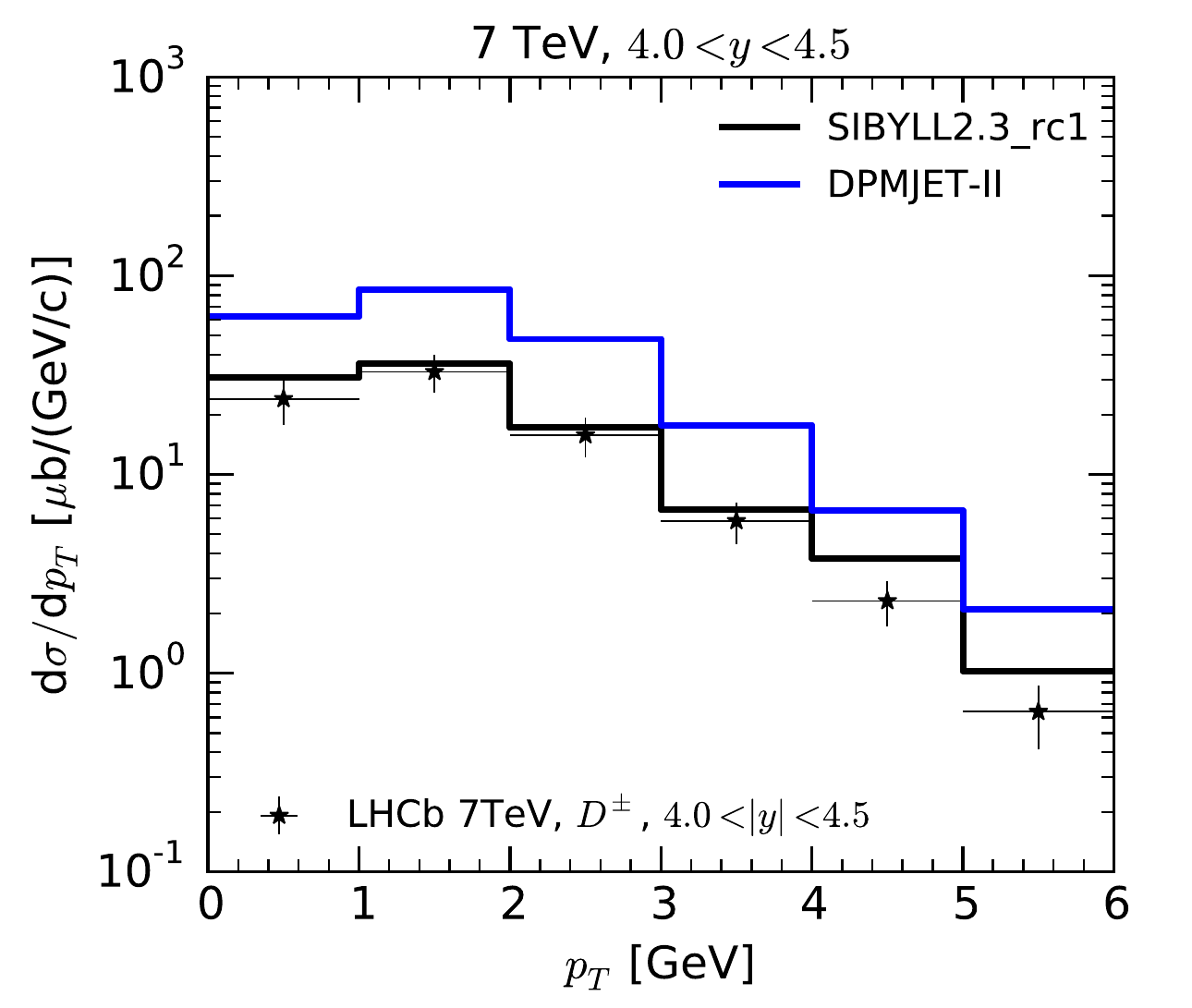}
  \caption{Comparison of the differential $D^\pm$ cross-section with data, measured in $pp$ collsions by LHCb \cite{TheLHCbCollaboration:2013uy}, with {\sc SIBYLL} and {\sc DPMJET} calculations.
  \label{fig:dpmjet_differential}}
\end{figure}

To examine the validity of charm production in DPMJET, we compare the total $c\bar{c}$ cross-sections with ALICE data in Fig.~\ref{fig:charm_models} and the differential $D^\pm$-meson cross-section with forward data from LHCb in Fig.~\ref{fig:dpmjet_differential}. The comparisons show that the charm model in {\sc DPMJET} overestimates the cross-sections in the forward phase-space. {\sc DPMJET} predictions are disfavored by LHC data. The recent measurements reduce the uncertainty of charm production for $pp$ collisions at PeV (Lab) energies. However, a larger fraction of uncertainty comes also from nuclear effects. To assess this uncertainty in a quantitative way, we make the assumption for the $c\bar{c}$ nuclear modification factor
\begin{equation}
  R_{p-air} = \frac{{\rm d}N^{c\bar{c}}_{p-air}/{\rm d}p_T}{\left<N_{coll}\right>
  {\rm d}N^{c\bar{c}}_{pp}/{\rm d}p_T} \equiv 1.
    \label{eq:scaling}
\end{equation}
In other words, there are no screening effects and the production of charmed quarks is a point-like process. Quantitatively this means for the inclusive $c\bar{c}$ cross-section in proton-air collisions
\begin{equation}
  \sigma_{c\bar{c}, p-air} = A_{air}\ \sigma_{c\bar{c}, p-p} = 14.5\ \sigma_{c\bar{c}, p-p}.
\end{equation}
In calculations labeled {\sc SIBYLL-2.3 PL} (PL for point-like) we use the charm cross sections and distributions of {\sc SIBYLL-2.3 RC1} as predicted for $pp$ interactions and scale the yields according to Eq.~\ref{eq:scaling}. 
\begin{figure}
  \centering
  \includegraphics[width=\columnwidth]{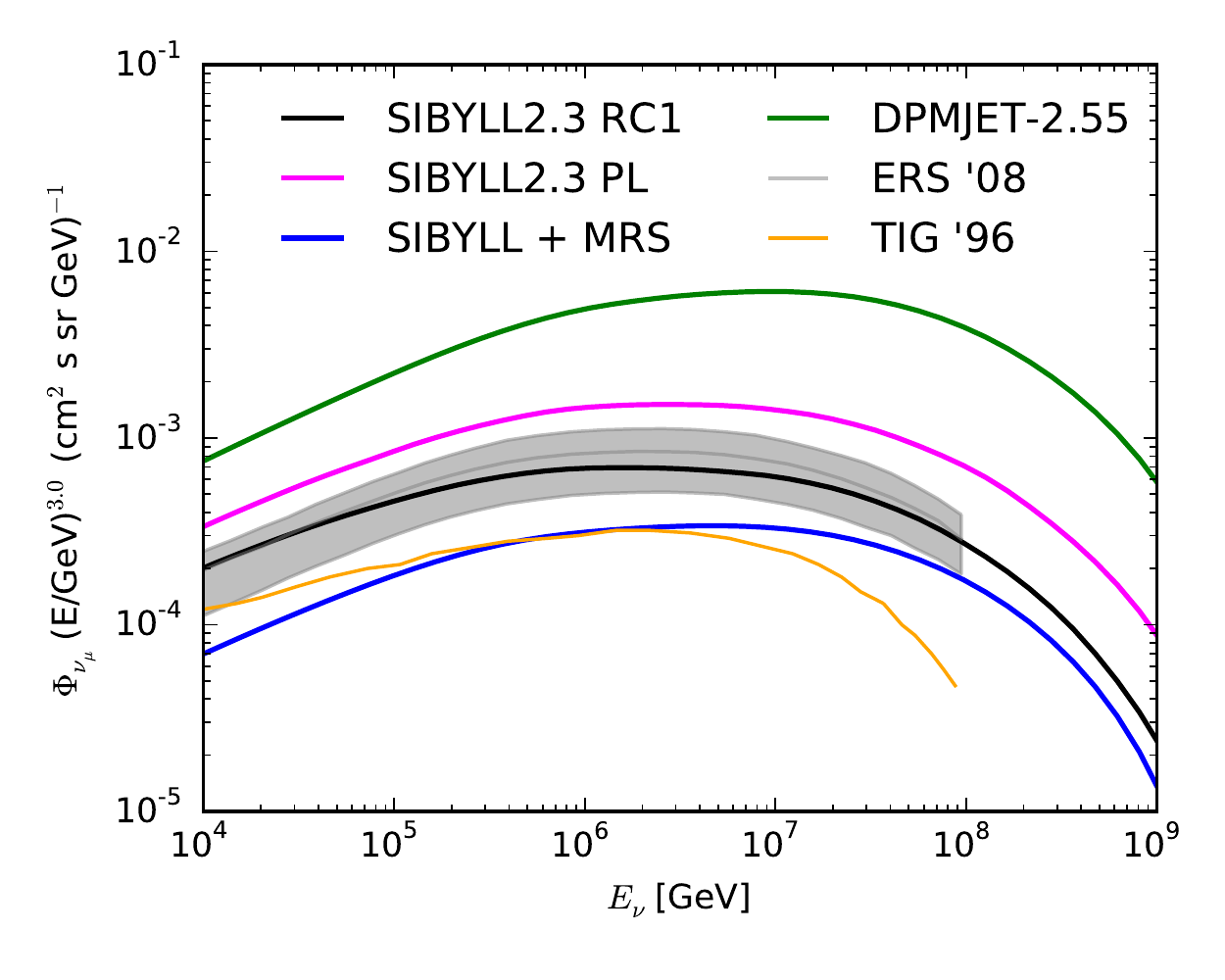}
  \caption{Prompt muon neutrino flux calculated using models described in the text. To allow for direct comparison, we use the same primary flux model as in ERS and TIG.
  \label{fig:prompt_models}}
\end{figure}

In Fig.~\ref{fig:prompt_models} the results of the different charm production models are compared with the Enberg-Reno-Sarcevic (ERS) \cite{enberg_2008} and the Thunman-Ingelman-Gondolo (TIG) \cite{thunman_1996} calculations of the prompt muon neutrino flux. {\sc SIBYLL-2.3 RC1} performs similarly to ERS and it produces notably higher fluxes than the MRS saturation model. All models predict different spectral shapes. Using {\sc DPMJET-II} results in an order of magnitude higher fluxes. We consider
the {\sc SIBYLL-2.3 PL} predictions as upper boundary for nuclear uncertainties of charm production.

\subsection{Partial contributions of intermediate particles}
\label{sec:intermediate_particles}
Since fluxes of all possible intermediate mesons and baryons are stored in the state vector $\Phi$, we can easily trace back the mother particles of leptons at the surface. 
\begin{figure*}
\centering
\includegraphics[width=\textwidth]{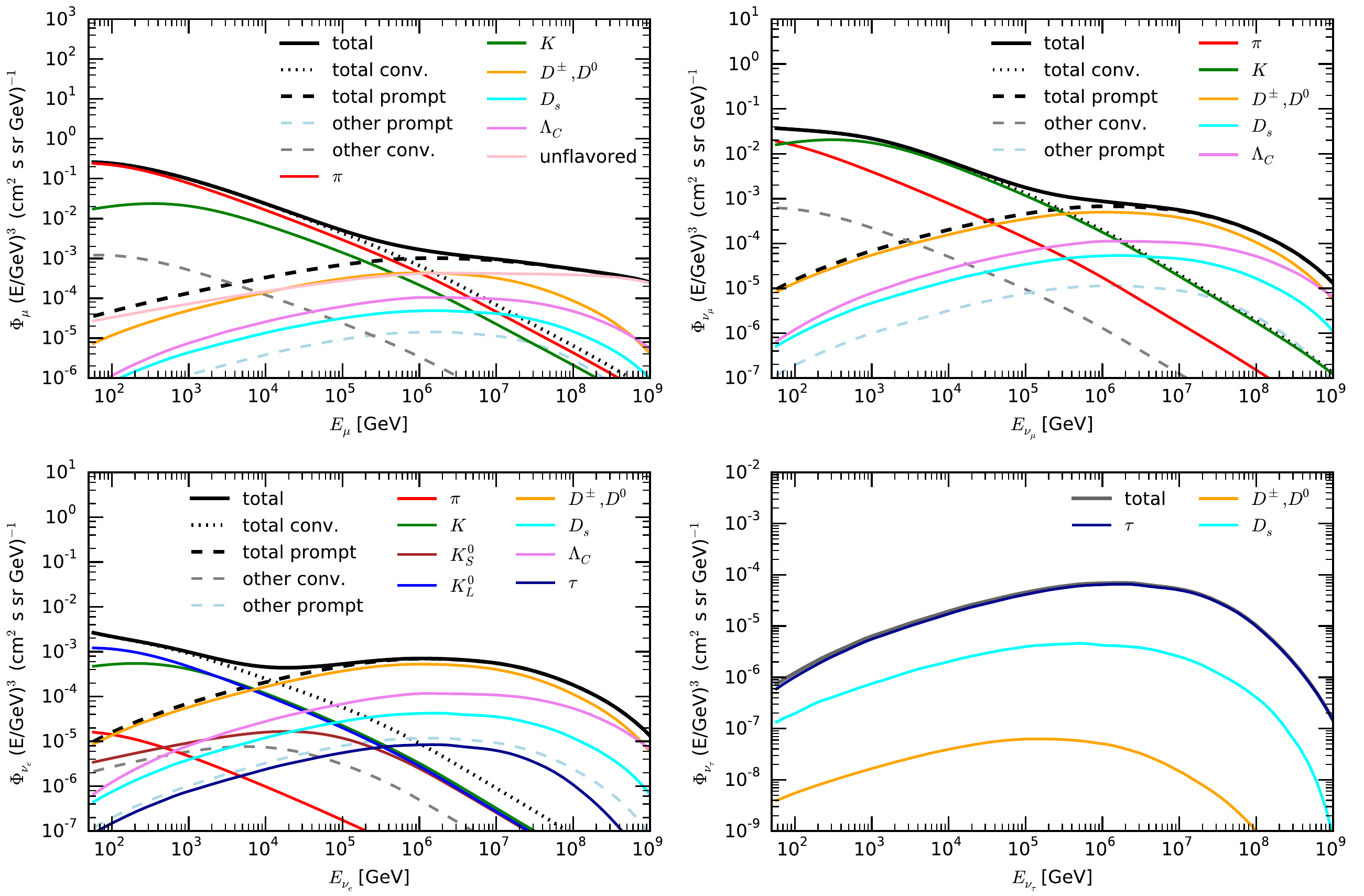}
\caption{Partial contribution of intermediate particles to the flux of atmospheric muons $\mu^+ + \mu^-$ (top left), muon neutrinos $\nu_\mu + \bar{\nu}_\mu$ (top right), electron neutrinos $\nu_e + \bar{\nu}_e$ (bottom left) and tau neutrinos $\nu_\tau + \bar{\nu}_\tau$ (bottom right). The primary spectrum is TIG and the interaction model is {\sc SIBYLL-2.3 RC1}.}
\label{fig:flux_details}       
\end{figure*}
Fig.~\ref{fig:flux_details} is a break down of the different contributions of intermediate particles to the conventional and prompt flux. We define a lepton as prompt, if its mother particle of the last decay has a $c\tau < c\tau(K^0_S) = 2.68$ cm. To improve the clarity of the graph, the simple broken power-law primary spectrum of TIG is employed as initial state. The dominant contributions to conventional muons are decays of charged pions and kaons, while prompt muons are originating from decays of charged and neutral $D$ mesons and unflavored mesons, such as $\eta$, $\omega$ and $\phi$. 
\begin{figure}
  \centering
  \includegraphics[width=\columnwidth]{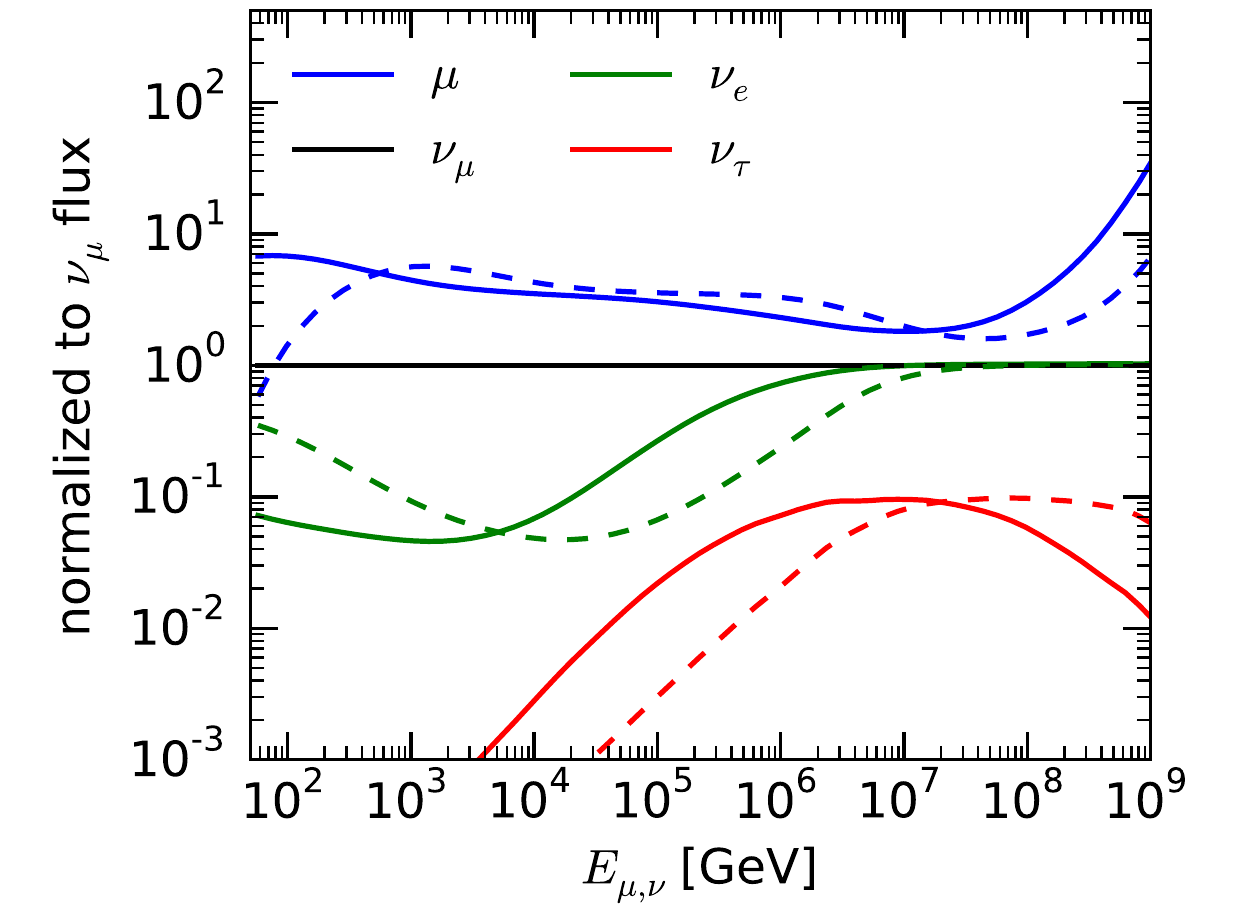}
  \caption{Flavor ratios of leptons at the surface, normalized to the muon neutrino flux. The calculation was performed using H3a primary flux and {\sc SIBYLL-2.3 RC1} for $\theta=0^\circ$ (solid) and $\theta=90^\circ$ (dashed).
  \label{fig:flavor_ratios}}
\end{figure}
The latter resonances break the correlation between muon and neutrino fluxes at very high energies~\cite{2009JCAP...09..008I} as shown in the flavor ratios of Fig.~\ref{fig:flavor_ratios}. Prompt muon neutrino and electron neutrino fluxes are roughly equal and originate from decays of $D$-mesons and $\Lambda_C^+$ baryons. The fractional contribution of $D$ and $\Lambda_C^+$ becomes equal at several hundreds of PeV. Decays of $K^\pm$ are responsible for the largest fraction of conventional muon neutrinos at energies above a few TeV. For electron neutrinos, other channels are important, such as decays of $K^0_L$. At several hundreds of TeV there is an additional contribution from $K^0_S$ as recently discussed in~\cite{2014arXiv1409.4924G}. In any case, the flux from charmed particles is expected to be significantly higher than that due to these other channels.

\section{Summary and Outlook}
\label{sec:summary}
An efficient numerical treatment of cascade equations has been developed for the calculation of atmospheric lepton fluxes at very high energy, with particular attention put on the transition from conventional to prompt production processes. As a first application we have shown calculations to illustrate the importance of the primary flux parametrization, the model of the atmosphere, the role of short-lived particles, and the model of charmed hadron production. More details about the numerical solution and the code will be published elsewhere.

\subsection*{Acknowledgments}
We gratefully acknowledge many inspiring and fruitful discussions
with colleagues of the IceCube Collaboration. 
This work supported by the Wolfgang-Gentner-Programme of the Bundesministerium für Bildung und Forschung (BMBF) and in part by the Helmholtz Alliance for Astroparticle
Physics HAP, which is funded by the Initiative and Networking Fund of
the Helmholtz Association. 

\bibliography{isvhecri2014.bib}
\end{document}